\documentclass{article}

\usepackage{PRIMEarxiv}

\usepackage[utf8]{inputenc} 
\usepackage[T1]{fontenc}    
\usepackage{hyperref}       
\usepackage{url}            
\usepackage{booktabs}       
\usepackage{amsfonts}       
\usepackage{nicefrac}       
\usepackage{microtype}      
\usepackage{lipsum}
\usepackage{fancyhdr}       
\usepackage{graphicx}       
\graphicspath{{media/}}     

\title{Deceptive AI systems that give explanations are just as convincing as honest AI systems in human-machine decision making
}

\author{
  Valdemar Danry \\
  MIT Media Lab \\
  Massachusetts Institute of Technology \\
  Cambridge \\
  \texttt{vdanry@mit.edu} \\
  \And
  Pat Pataranutaporn \\
  MIT Media Lab \\
  Massachusetts Institute of Technology \\
  Cambridge \\
  \texttt{patpat@media.mit.edu} \\
  \And
  Ziv Epstein \\
  MIT Media Lab \\
  Massachusetts Institute of Technology \\
  Cambridge \\
  \texttt{zive@media.mit.edu} \\
  \And
  Matthew Groh \\
  MIT Media Lab \\
  Massachusetts Institute of Technology \\
  Cambridge \\
  \texttt{groh@media.mit.edu} \\
  \And
  Pattie Maes \\
  MIT Media Lab \\
  Massachusetts Institute of Technology \\
  Cambridge \\
  \texttt{pattie@media.mit.edu} \\
}

\begin{document}
\maketitle



\section*{Extended Abstract}
In the past few years, there has been an increase in AI-based disinformation campaigns, which are attempts to spread misinformation online for strategic reasons \cite{kertysova2018artificial}. How AI-systems explain how they arrive at their classifications can be deceptive, in that they can be manipulated to make the system appear more reliable than it is \cite{lakkaraju2020fool}. For example, a bot may claim to be human in order to evade detection, or a machine learning system may falsely claim a piece of information to be true when it is not. While previous work has shown that AI-explanations help people determine the veracity of information online and change people’s beliefs \cite{lai2019human}, little is known about how susceptible people are to deceptive AI systems. For instance, previous research on placebic information has shown that any explanation (even poor ones) significantly influences people’s behavior \cite{langer1978mindlessness}. With the increasing prevalence of large language models like GPT-3 that can automatically generate and target individuals with highly believable and deceptive explanations to manipulate their opinion \cite{brown2020language}, and with the same models increasingly being proposed in AI fact-checking systems \cite{stammbach2020fever}, a practical question rooted in the theory of placebic information emerges: how do AI systems with honest and deceptive explanations affect people's ability to discern true news from fake news online?

In this paper, we investigate how people's discernment varies (1) when AI systems are perceived as either human fact-checkers or AI fact-checking systems, and (2) when explanations provided by those fact-checkers are either deceptive (i.e. the AI system falsely generating explanations for why a true headline is false or why a false headline is true) or honest (i.e. the AI system accurately generating explanations for why a true headline is true or why a false headline is false). In a between-subjects randomized 2×2 factorial design experiment, we had 128 participants provide 1,792 truth discernment judgments on 14 different true and false news headlines. The headlines were randomly assigned with AI-generated explanations that (1) are either labeled as a “human fact-checker” or an “AI fact-checking system”, and (2) are without the participants knowledge being either deceptive or honest. Participants were asked to provide judgment discerning true news from fake news on a likert scale (from “definitely false” to “definitely true”), which is used to calculate weighted discernment accuracy. Participants were asked to provide their judgment twice, one before the explanation (pre-explanation), and one after the explanation, also subjected to self reporting their level of trust in the agent providing them with explanations.

We created a dataset of headlines each with one honest and one deceptive explanation by prompting the state-of-the-art text-generation model GPT-3 (175 billion hyperparameters) with 12 example explanations randomly sampled from the publicly available fact-checking dataset “liar-plus” \cite{alhindi2018your}. This dataset consists of 12,836 short statements with explanation sentences extracted automatically from the full-text verdict reports written by journalists in Politifact. First, 5 honest and 5 deceptive explanations were generated for 40 true and false headlines by prompting GPT-3 (davinci, \textit{temp = .7}) with the headline and making it complete the sentences “This is FALSE because…” or “This is TRUE because…” We further curated these explanations by ranking them by highest semantic similarity and lowest repeated-word frequency. We picked the highest ranked explanations, confirmed the veracity and logical validity of each explanation, and randomly excluded generated explanations until we had an equal distribution of true and false explanations for each condition (deceptive vs. honest explanations and AI fact-checker vs. human fact-checker) ending with a stimulus set consisting of 14 headlines with 1 honest explanation and 1 deceptive explanation each. We tested for differences across four linguistic dimensions (wordcount, sentiment, grade level, and subjectivity) and found no statistical differences between conditions. 

\begin{figure}
  \centering
  \includegraphics[width=1\textwidth]{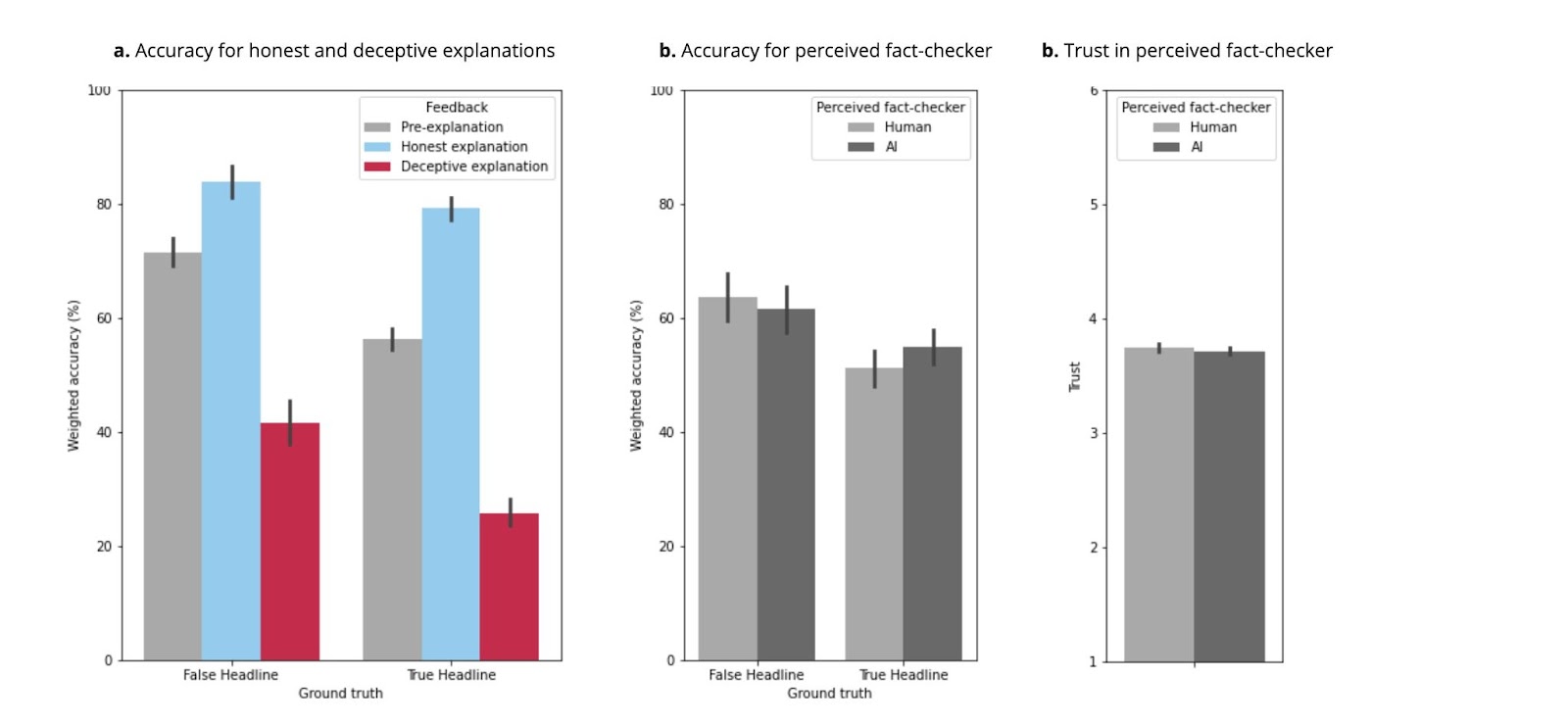}
  \caption{Figure 1. The mean and distribution of (a) participants’ weighted accuracy for honest and deceptive explanations, (b) weighted accuracy with AI explanations perceived as coming from a human or AI fact-checker, cd) participant trust in human or AI fact-checkers. The black lines on each box indicate the 95\% confidence interval of the true mean, while connecting lines between boxes indicate the significant difference between the means of two groups.}
  \label{fig:fig1}
\end{figure}

A two sample t-test was performed to (1) compare the impact of deceptive and honest explanations on weighted discernment accuracy of headlines, and (2) to compare weighted discernment accuracy when the AI explanations are perceived as coming from either a human or AI fact-checker. We find that deceitful explanations significantly reduce accuracy, indicating that people are just as likely to believe deceptive AI explanations as honest AI explanations (see figure \ref{fig:fig1}). Although before getting assistance from an AI-system (pre-explanation), people have significantly higher weighted discernment accuracy on false headlines (\textit{mean = 62.7\%, SD = 37.9\%}) than true headlines (\textit{mean = 53.1\%, SD = 36.2\%}), \textit{t = 9.87, p = .000}, we found that with assistance from an AI system, discernment accuracy increased significantly when given honest explanations on both true headlines (\textit{mean = 83.9, SD = 25.6}), \textit{t = -6.12, p = .000} and false headlines  (\textit{mean = 79.3, SD = 22.2 }), \textit{t = -15.3, p = .000}, and decreased significantly when given deceitful explanations on true headlines (\textit{mean = 41.7, SD = 36.4}), \textit{t =13.32, p = .000}, and false headlines (\textit{mean = 25.79, SD = 26.4}), \textit{t = 19.1, p = .000}. Further, we did not observe any significant differences in discernment between explanations perceived as coming from a human fact checker (\textit{mean = 56.1\%, SD = 37.9\%}) compared to an AI-fact checker (\textit{mean = 57.6\%, SD = 36.5\%}), \textit{t = -0.79,  p = .427}. Similarly, we found no significant differences in trust between human fact-checkers (\textit{mean = 3.74, SD = 0.44}) and AI fact-checkers (\textit{mean = 3.72, SD = 3.72}), \textit{t = 1.47, p = .141}.

\bibliographystyle{unsrt}  
\bibliography{references}

\end{document}